\documentstyle[aps,prl,preprint,floats,epsfig,lscape]{revtex}

\begin{document}

\preprint{\tighten\vbox{\hbox{\hfil CLNS 02/1792}
                        \hbox{\hfil CLEO 02-10}
}}

\title{Dalitz Analysis of $D^0 \to K_S^0\pi^+\pi^-$}  

\author{CLEO Collaboration}
\date{July 22, 2002}

\maketitle
\tighten

\begin{abstract} 
In $e^+e^-$ collisions using the CLEO detector we have studied the decay
of the $D^0$ to the final state $K_S^0\pi^+\pi^-$ with the initial
flavor of the $D^0$ tagged in charged $D^\ast$ decay.  We use the Dalitz
technique to measure the resonant substructure in this final state and clearly
observe ten different contributions by fitting for their amplitudes
and relative phases.
We observe a $K^{\ast +} \pi^-$ component which arises from doubly Cabibbo
suppressed decays or $D^0-\overline{D^0}$ mixing.
\end{abstract}

\newpage

{
\renewcommand{\thefootnote}{\fnsymbol{footnote}}

\begin{center}
H.~Muramatsu,$^{1}$ S.~J.~Richichi,$^{1}$ H.~Severini,$^{1}$
P.~Skubic,$^{1}$
S.A.~Dytman,$^{2}$ J.A.~Mueller,$^{2}$ S.~Nam,$^{2}$
V.~Savinov,$^{2}$
S.~Chen,$^{3}$ J.~W.~Hinson,$^{3}$
% ...
\end{center}
 
\small
\begin{center}
$^{1}${University of Oklahoma, Norman, Oklahoma 73019}\\
$^{2}${University of Pittsburgh, Pittsburgh, Pennsylvania 15260}\\
$^{3}${Purdue University, West Lafayette, Indiana 47907}\\
% ...
\end{center}

\setcounter{footnote}{0}
}
\newpage

Weak hadronic decays of charmed mesons are expected to proceed
dominantly by resonant two-body decays in several theoretical models
~\cite{ref:bsw,ref:bedaque,ref:chau,ref:terasaki,ref:buccella}.
A clearer understanding of final state interactions in exclusive weak
decays is an important ingredient for our ability to model
decay rates as well as for our understanding of interesting phenomena
such as mixing~\cite{ref:hepph9802291}.
In this context an interesting final state is $D^0 \to \overline{K}^0\pi^+\pi^-$ which can proceed through a number of two-body states.
Previous investigations~\cite{MarkIII,e687a,E691,argus,e687b} of the 
substructure in this channel were limited by statistics to the Cabibbo
favored decays.  
A key motivation is to observe one or more of the $D^0 \to K^0\pi^+\pi^-$
resonant submodes that proceed via mixing or double Cabibbo suppression,
such as $K^{\ast+}\pi^-$ or ${K}_{0}\!(1430)^+\pi^-$, and to measure their
phase relative to the corresponding unsuppressed $\overline{K}^0\pi^+\pi^-$
submodes.

This analysis uses an integrated luminosity of 9.0~fb$^{-1}$
of $e^+e^-$ collisions at $\sqrt{s}\approx10\,$GeV provided by
the Cornell Electron Storage Ring (CESR).
The data were taken with the CLEO~II.V configuration of the CLEO~II
multipurpose detector~\cite{ctwo}. A silicon vertex detector (SVX) 
was installed in the upgraded configuration~\cite{csvx}.

The event selection is similar to that used in our search for 
$D^0$-$\overline{D^0}$ mixing via the process $D^0 \rightarrow
\overline{D^0} \rightarrow K^+ \pi^-$~\cite{kpi}.
We reconstruct candidates for the decay sequence
$D^{\ast+}\!\to\!\pi^+_s D^0$, $D^0\!\to\!K^0_S\pi^+\pi^-$.
Consideration of charge conjugated modes is implied
throughout this Letter.
The charge of the slow
pion ($\pi^+_s$ or $\pi^-_s$) identifies the charm state 
at $t=0$ as either $D^0$ or $\overline{D^0}$. 
We require the $D^{\ast+}$ momentum
$p_{D^\ast}$ to exceed 2.0 GeV$/c$, and we require the
$D^0$ to produce the final state
$K_S^0\pi^+\pi^-$. We reconstruct $K^0_S\!\to\!\pi^+\pi^-$ with the
requirement that the daughter pion tracks form a common vertex, 
in three dimensions, with a confidence level $>10^{-6}$.
Signal candidates pass
the vertex requirement with $96\%$ relative efficiency. Throughout this Letter,
relative efficiency is defined as the number of events passing all 
requirements relative to the number of events when only the
requirement under
study is relaxed.

Our silicon vertex detector provides precise measurement of the charged 
tracks in three dimensions~\cite{svxres}.
We exploit the precision tracking of the SVX by refitting the 
$K_S^0$ and $\pi^\pm$ tracks with a requirement that they
form a common vertex in three dimensions.
We use the trajectory of the
$K^0_S\pi^+\pi^-$ system and the position of the CESR luminous region
to obtain the $D^0$ production point.  
We refit the $\pi_s^+$ track with a requirement that
the trajectory intersect
the $D^0$ production point.
We require that the
confidence level of each refit exceed $10^{-4}$. 
The signal candidates pass the $D^0$ mass and decay vertex requirement 
with $85\%$ and $91\%$ relative efficiency, respectively. 
                  
We reconstruct the energy released in the 
$D^{\ast+}\!\to\!\pi^+_sD^0$ decay as $Q\!\equiv\!M^\ast\!-\!M\!-\!m_\pi$, 
where $M^\ast$ is the reconstructed mass of the
$\pi_s^+ K^0_S \pi^+ \pi^-$ system, $M$ is the reconstructed mass of the
$K^0_S \pi^+\pi^-$ system, and $m_\pi$ is the charged pion mass.  
The addition of the $D^0$ production point to the $\pi_s^+$
trajectory
yields the resolution $\sigma_Q = 220\pm$4~keV,
where $\sigma_Q$ is the core value from a fit to a bifurcated student's $t$
distribution.
We obtain a resolution on $M$ of
$\sigma_M = 4.8\pm0.1\,$MeV and a resolution on
$m_{K^0_S}$ of $\sigma_{m_{K^0_S}} =2.4\pm0.1\,$MeV. 

\begin{figure}
\begin{center}
\epsfig{figure=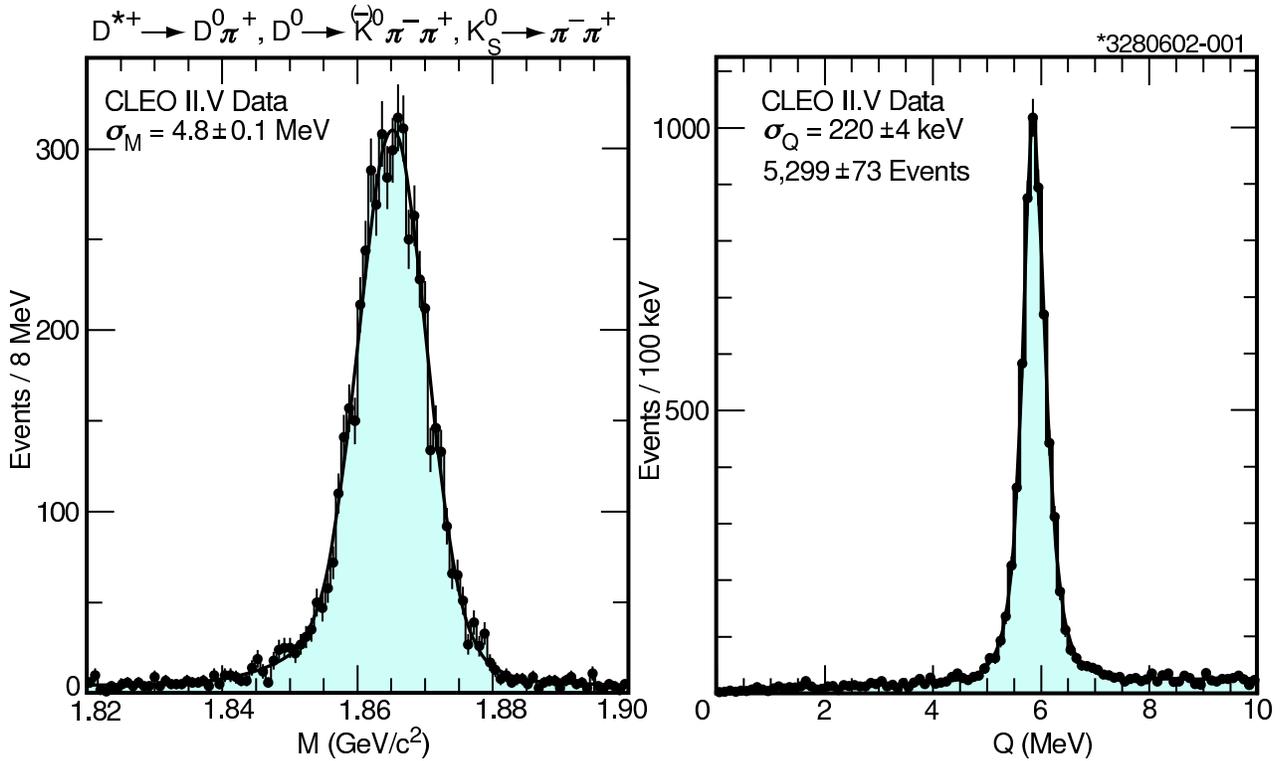}
\end{center}
\caption{Distribution of a)
$Q$ and b) $M$
for the process $D^0 \to K^0_S\pi^+\pi^-$.
The candidates  
pass all selection criteria discussed in the text.}
\label{fig:qandm}
\end{figure}

We apply a
set of `prophylactic' requirements 
to exclude candidates with a poorly determined 
$Q$, $M$, or $D^0$ flight time, and $K^0_S$
candidates that are likely to be background.
The typical computed $\sigma$ for $Q$, $M$ and flight time
error is $150$~keV, $5$~MeV and $0.5~\tau_{D^0}$, respectively.
These are computed from the reconstruction covariance matrix of the 
daughters of the $D^0$ candidate. 
We reject
candidates where $\sigma_Q$, $\sigma_M$ or 
flight time error exceeds 300~keV, 10~MeV or 2.0$\tau_{D^0}$, respectively.
The relative efficiencies for the 
signal candidates to pass these cuts is 98\%, 98\% and 89\%, respectively. 
We exclude $K^0_S$ candidates with a vertical flight distance less then $500$ 
microns to remove combinatoric background with zero lifetime. The signal
candidates survive this requirement with $95\%$ relative efficiency.
The distributions of $Q$ and $M$ for our data are shown in
Figure~\ref{fig:qandm}.

	We select 5299 candidates within three standard deviations of
the expected $Q$, $M$, and $m_{K_S^0}$.  The efficiency for the selection
described above is nearly uniform across the Dalitz distribution.
In our simulation we generate the $D^0 \to K_S^0\pi^+\pi^-$ 
uniformly populating the allowed phase space.  
We study our efficiency with a GEANT~\cite{GEANT} based simulation of
$e^+ e^- \to c \overline c$ events in our detector with a luminosity corresponding
to more than three times our data sample.  
We observe that our selection introduces distortions due
to inefficiencies near the edge of phase space, and fit the efficiency to a two
dimensional cubic
polynomial in ($m_{K_S^0\pi^-}^2$,$m_{\pi^+\pi^-}^2$) requiring that the
efficiency does not change if the $\pi^+$ and $\pi^-$ are interchanged.
Our standard result uses this efficiency
parameterization to interpret the Dalitz distribution.  To take
into account a systematic uncertainty in our selection efficiency we
compare the standard result with an efficiency that is uniform across the
allowed Dalitz distribution.

	Figure~\ref{fig:qandm} shows that the background is small, but non-negligible.
Fitting the $M$ distribution to a signal shape as described
above plus a quadratic
background shape we find a background fraction of $2.1 \pm 1.5$\%.  We use
this fraction as a constraint when fitting the Dalitz distribution.  To
model the background contribution in the Dalitz distribution we consider those
events in the data that are in sidebands five to ten standard deviations
from the signal in $Q$ and $M$ and within three in $m_{K^0_S}$.
There are 445 candidates in this selection, about four times the amount
of background we estimate from the signal region.  We compare these with
background from our simulation also including $e^+e^-$ annihilations producing
the lighter quarks.  We note that the background from the simulation is 
dominated by random combinations of unrelated tracks, and the shape of
the background in the simulation and the sideband data sample
agree well.  The simulation predicts that the background uniformly populates
the allowed phase space, and we model this contribution to the Dalitz
distribution by fitting the data sideband sample to a two dimensional
cubic polynomial in ($m_{K_S^0\pi^-}^2$,$m_{\pi^+\pi^-}^2$).
All parameters except the constant are
consistent with zero.
Other possible contributions to the background, resonances
combined with random tracks and real $D^0$ decays combined with random
soft pions of the wrong charge, are negligible in the simulation.  The latter,
called mistags,
are especially dangerous to our search for ``wrong sign'' $D^0$ decays.
Mistags populate the Dalitz distribution in a known way that depends on the
shape of signal.  When we analyze the Dalitz distribution we allow a mistag
fraction with an unconstrained contribution.  We have looked for the
contribution of an anomalous resonance, such as $\rho^0$ or
$K^{\ast-}$, plus random tracks to the background in the data
by studying the sidebands in $Q$, $M$, and $m_{K^0_S}$, and
conclude that any such contributions are negligible.

\begin{figure}[t]
\begin{center}
\epsfig{figure=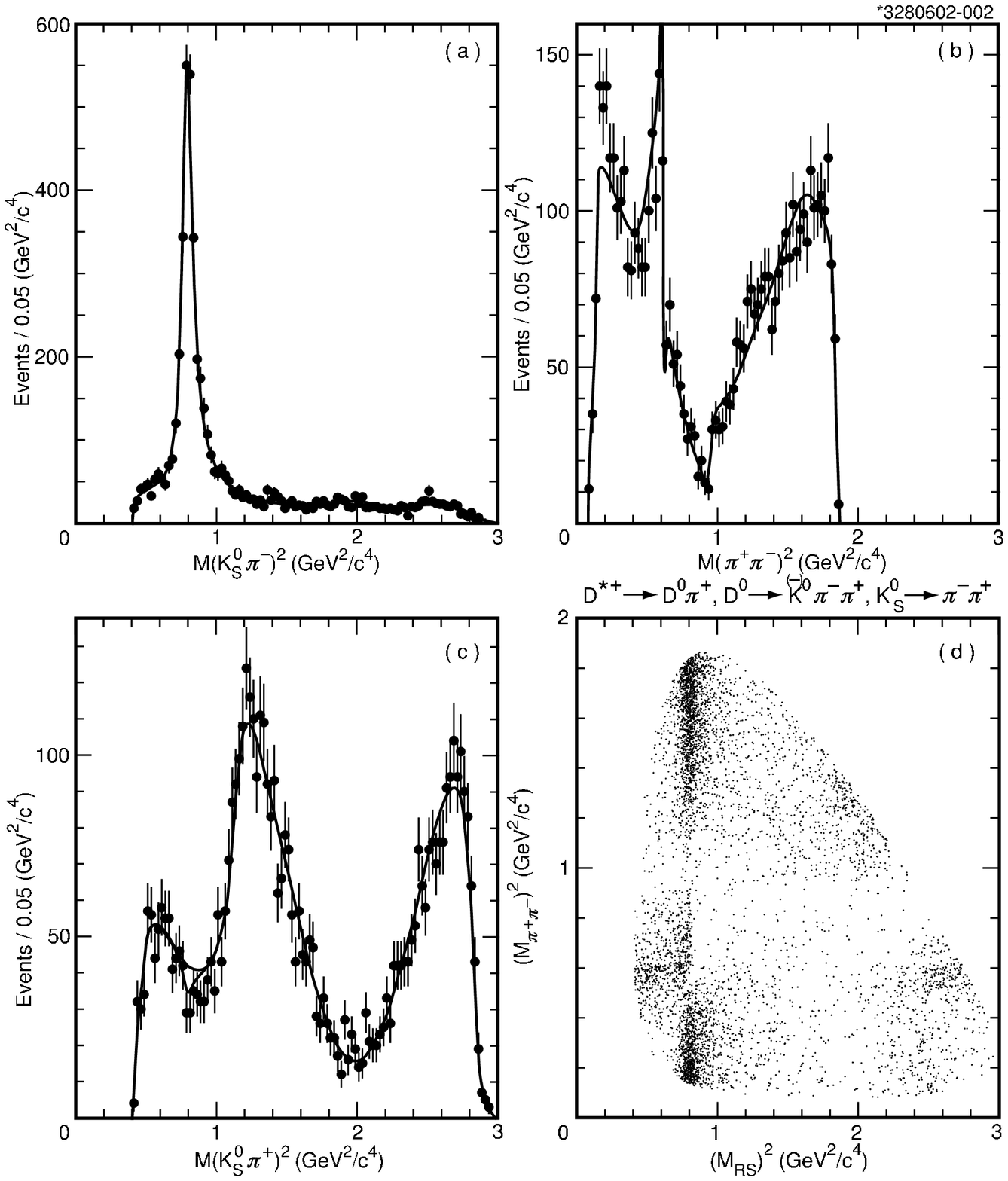}
\end{center}
\caption
{\label{fig:proj}
Projections of the results of the fit described in the text
to the $K^0_S\pi^+\pi^-$~Dalitz distribution showing both the
fit (histogram) and the data (points).  In c) the result of a
fit where the ``wrong sign'' $D^0 \to K^{\ast+} \pi^-$ amplitude
is fixed to zero is also shown.
d) The Dalitz distribution for $D^0\to K^0_S\pi^+\pi^-$ candidates. 
The horizontal
axis $(M_{RS})^2$ corresponds to $(M_{K^0_S\pi^-})$ for $D^0$ and 
$(M_{K^0_S\pi^+})$ for $\overline{D^0}$.}
\end{figure}

	
Figure~\ref{fig:proj} shows the Dalitz distribution for the
$D^0 \to K_S^0\pi^+\pi^-$
candidates.  A rich structure is evident.  Contributions from
$K^{\ast-}\pi^+$ and $K_S^0\rho^0$ are apparent. Depopulated
regions exist suggesting destructive interference between some resonances and the
dominant decay modes.

We parameterize the $K^0_S\pi^+\pi^-$ Dalitz distribution following 
the methodology described in Ref.~\cite{bergfeld} using the same sign
convention used in previous investigations of this
decay channel~\cite{e687a,e687b}.  
We consider nineteen resonant subcomponents, 
$K^{\ast-} \pi^+$, 
$\kappa(800)^- \pi^+$,
${K}^{\ast}\!(1410)^- \pi^+$,
${K}^{\ast}_{0}\!(1430)^- \pi^+$,
${K}^{\ast}_{2}\!(1430)^- \pi^+$,
${K}^{\ast}(1680)^- \pi^+$,
${K^{\ast}_3}(1780)^- \pi^+$,
$K^0_S \rho$,
$K^0_S \omega$,
$K^0_S \rho(1450)$,
$K^0_S \rho(1700)$,
$K^0_S \sigma(500)$, 
$K^0_S f_0(980)$,
$K^0_S f_2(1270)$, 
$K^0_S f_0(1370)$, 
$K^0_S f_0(1500)$, 
$K^0_S f_0(1710)$, and the
``wrong sign'' $K^{\ast+} \pi^-$
and ${K}^{\ast}_{0}\!(1430)^+ \pi^-$,
as well as a non-resonant contribution.
The parameters of the resonances are taken from Ref.~\cite{pdg} except
for the $f_0$'s. We use Ref.~\cite{Aitala:2000xt}
for the $f_0(980)$ and the coupled channel analysis of
Ref.~\cite{Kirk:2000ay} for the $f_0(1370)$, $f_0(1500)$ and
$f_0(1710)$.
We consider that each of the resonances has its own amplitude
and relative phase.  The non-resonant contribution is 
modeled as a uniform distribution across the allowed phase space with
a fixed relative phase.  The phases and widths of the resonance contributions
vary as given by the spin of the resonance as described in Ref.~\cite{bergfeld}.


This study is sensitive only to relative phases 
and amplitudes.
Thus we fix one phase and one
amplitude.  To minimize correlated errors on the phases and
amplitudes we choose the largest color suppressed mode,
$K^0_S \rho$, which should be out of phase with the
unsuppressed modes to have a fixed zero phase and an amplitude of one.
Since the choice of normalization, phase convention, and amplitude
formalism may not always be identical for different experiments, 
fit fractions are reported in addition to amplitudes to allow for more meaningful 
comparisons between results. 
The fit fraction is defined as the integral of a single
component divided by the coherent sum of all components.
The sum of the fit fractions for all components
will in general not be unity because of the effect of interference.

	Backgrounds, combinatorics and mistags,
are considered as described above.  They do
not interfere with the signal, but the mistag background shape
depends on the signal shape as noted earlier.

One must also consider the statistical errors on the fit fractions. We have
chosen to use the full covariance matrix from the fits to determine
the errors on fit fractions so that the assigned errors will properly
include the correlated components of the errors on the amplitudes and
phases. After each fit, the covariance matrix and final parameter
values are used to generate 500 sample parameter sets.  For each set,
the fit fractions are calculated and recorded in histograms. Each
histogram is fit with a single Gaussian to extract its width, which
is used as a measure of the statistical error on the fit fraction.

	We perform an initial fit,
using the unbinned maximum likelihood technique
including the resonances observed by E687~\cite{e687b}.  
We then consider each of the intermediates states listed above
retaining those that are more than three standard deviations significant.
We do not find the $\kappa(800)^- \pi^+$, ${K}^{\ast}\!(1410)^- \pi^+$,
${K^{\ast}_3}(1780)^- \pi^+$, $K^0_S \rho(1450)$, $K^0_S \rho(1700)$,
$K^0_S f_0(1500)$, $K^0_S f_0(1710)$,
and the``wrong sign'' ${K}^{\ast}_{0}\!(1430)^+ \pi^-$ to be significant.
The $K^0_S \sigma(500)$ is a special case.  It is excluded
in our standard fit, and its possible contribution is discussed further below.
The remaining ten resonances,
a non-resonant contribution, and backgrounds as described above are
included in our standard fit and give our central results.  

	Table \ref{tab:fit} gives the results of our standard fit. 
Figure~\ref{fig:proj} shows the three projections of the fit.  
We note that there is a significant ``wrong sign''
$D^0 \to K^{\ast+} \pi^-$ amplitude.  
Mistags are not significant, having a rate of $0.1\pm0.4$\%.
When we compare the likelihood of our
standard fit to one where the $K^{\ast+} \pi^-$ amplitude fixed to zero we see that
the statistical significance of the $K^{\ast+} \pi^-$ amplitude
is 5.5 standard deviations.  Also we note that the phase difference
between the $K^{\ast-} \pi^+$ and $K^{\ast+} \pi^-$ contributions 
is consistent with $180^\circ$, as expected from Cabibbo factors.

	We consider systematic uncertainties that arise from our 
model of the background, the efficiency, and biases due to experimental
resolution.  
Our general procedure is to change some aspect of our standard fit and
interpret the change in the values of the amplitudes and phases as an
estimate of the systematic uncertainty.
The background is modeled with a two dimensional cubic polynomial and
the covariance matrix of the polynomial coefficents determined
from a sideband.  Our standard fit fixes the coefficients of the background
polynomial, and to estimate the systematic uncertainty on 
this background shape we perform a fit with the coefficients allowed
to float constrained by the covariance matrix.
Similarly we perform a fit with a uniform efficiency rather than
the efficiency shape determined from the simulation as an estimate
of the systematic uncertainty due to the efficiency. 
We change selection criteria in the
analysis to test whether our simulation properly models the efficiency.
These variations to the standard fit are the largest contribution to
our experimental systematic errors.
To study the effect of the finite resolution our experiment has on the
variables in the Dalitz plots we vary the size of the bins used to compute the
overall normalization.

	Another class of systematic uncertainties arise from our choices
for the decay model for $D^0 \to K_S^0\pi^+\pi^-$.  We consider the
Zemach formalism \cite{zemach}, rather than
the standard helicity model, which enforces the transversality of 
intermediate resonances and we vary the radius parameter~\cite{blatweis} for
the intermediate resonances 
and for the $D^0$ between zero and twice
their standard value of $0.3\ {\rm fm}$ and $1\ {\rm fm}$, respectively.
These variations to the standard fit are the largest contribution to
our modeling systematic errors.
Additionally, we allow the masses and widths for the intermediate resonances to
vary within their known errors~\cite{pdg,Aitala:2000xt,Kirk:2000ay}.  

	We also consider uncertainty
arising from which resonances we choose to include in our fit to the Dalitz
plot.  
We compared the result of our standard fit to a series of fits where each of
the possible resonances were included one at a time. We also
considered a fit including all possible resonances.
We take the 
maximum variation of the
amplitudes and phases from the standard result compared to the results in
this series of fits as a measure of the uncertainty due to our choice
of included resonances.  

The $\sigma(500)$ has been reported by E791
in $D^+ \to \pi^-\pi^+\pi^+$ decays~\cite{sigma500}.  
The parameters of the $\sigma(500)$ are sensitive to the choice of decay model,
discussed previously.
Replacing the non-resonant contribution in our standard fit with a $K^0_S
\sigma(500)$ component yields an amplitude
of $0.57\pm0.13$ and phase of $214\pm 11$
with a mass of $m_\sigma= 513 \pm 32$ and width
$\Gamma_\sigma = 335 \pm 67$~MeV for the $\sigma(500)$
consistent with E791
results~\cite{sigma500} ($m_\sigma= 478 \pm 29$ and 
$\Gamma_\sigma = 324 \pm 46$~MeV).
While we find this suggestive that 
there is a $K^0_S \sigma(500)$ contribution, we
are unable to definitively confirm this because of the known shortcomings of
our description of the scalar resonances.
The systematic uncertainty does
include the difference between the standard fit
which does not include the $\sigma(500)$, and the fits allowing it. 
On the other hand, we find no evidence for
a scalar $\kappa^- \to K^0_S \pi^-$ as also suggested by E791~\cite{kappa800}
in charm decays.


	We also do separate standard fits for $D^0$ and
$\overline{D^0}$ tags to search
for $CP$ violating effects.  We see no statistically significant
difference between these two fits. A more general study
would consider a $CP$ violating amplitude for each component observed in
our standard fit.

	In conclusion, we have analyzed the resonant substructure of 
the decay $D^0 \to K_S^0\pi^+\pi^-$
using the Dalitz technique.
We observe ten contributions including a ``wrong sign''
$D^0 \to K^{\ast+} \pi^-$ amplitude with a significance of 5.5 standard
deviations.  This decay arises from a double Cabibbo suppressed
decay or $D^0-\overline{D^0}$ mixing.  
We measure
$\frac{{\cal B}(D^0 \to K^{\ast+} \pi^-)}{{\cal B}(D^0 \to
K^{\ast-} \pi^+)} = (0.5 \pm 0.2 ^{+0.5}_{-0.1}\,^{+0.4}_{-0.1})\%$,
and the relative phase between the two decays to be $(189\pm 10 \pm
3^{+15}_{-\,5})^\circ$, consistent with $180^\circ$.
We consider $D^0$ and
$\overline{D^0}$ tags separately,
and see no $CP$ violating effects.

	The time dependence of the Dalitz distribution of this decay mode
in our data is able to discern the source of the
``wrong sign'' signal, doubly Cabibbo suppressed decays or
$D^0-\overline{D^0}$ mixing, 
and have sensitivity to the mixing parameters $x$, explicitly to its
sign, and $y$ at the few percent level~\cite{asnernote}.

\section*{Acknowledgment}

We gratefully acknowledge the effort of the CESR staff in providing us with
excellent luminosity and running conditions.
This work was supported by 
the National Science Foundation,
the U.S. Department of Energy,
the Research Corporation,
and the Texas Advanced Research Program.


\begin{landscape}
\begin{table}[b] 
\caption{
Standard fit results. The errors shown are
statistical, experimental systematic, and modeling systematic
respectively.  See the text for further discussion.
}
\label{tab:fit}
\begin{tabular}{|l|ccc|}
Component     & Amplitude & 
Phase &
Fit Fraction (\%) \\ \hline
$K^\ast(892)^+ \pi^-$ $\times B(K^\ast(892)^+ \to K^0 \pi^+)$ &$(11 \pm 2\,\,^{+4}_{-1}\,\, ^{+4}_{-1}) \times 10^{-2}$&$321 \pm 
10 \pm 3\,\,^{+15}_{-5}$ &
$0.34 \pm 0.13\,\,^{+0.31}_{-0.03} \,\,^{+0.26}_{-0.02}$ \\
$\overline{K}^0 \rho^0$  & $ 1.0$ (fixed)   &
$0$ (fixed) & $26.4\pm0.9\,\,^{+0.9}_{-0.7} \,\,^{+0.4}_{-2.5}$ \\
$\overline{K}^0 \omega$ $\times B(\omega \to \pi^+ \pi^-)$ & $(37 \pm 5 \pm 1\,\,^{+3}_{-8}) \times 10^{-3}$&
$114 \pm 7\,\,^{+6}_{-4} \,\,^{+2}_{-5}$ &
$0.72 \pm 0.18\,\,^{+0.04}_{-0.06} \,\,^{+0.10}_{-0.07}$ \\ 
$K^\ast(892)^- \pi^+$ $\times B(K^\ast(892)^- \to {\overline K}^0 \pi^-)$  & $ 1.56\pm0.03 \pm0.02\,\,^{+0.15}_{-0.03}$ &
$150\pm 2\pm 2\,\,^{+2}_{-5}$  & 
$65.7\pm1.3 \,\,^{+1.1}_{-2.6}\,\,^{+1.4}_{-3.0} $ \\
$\overline{K}^0 f_0(980)$ $\times B(f_0(980) \to \pi^+ \pi^-)$ & $0.34 \pm 0.02\,\,^{+0.04}_{-0.03}\,\,^{+0.04}_{-0.02}$ &
$188 \pm 4\,\,^{+5}_{-3}\,\,^{+8}_{-6}$ &
$4.3 \pm 0.5\,\,^{+1.1}_{-0.4}\pm 0.5$ \\ 
$\overline{K}^0 f_2(1270)$ $\times B(f_2(1270) \to \pi^+ \pi^-)$ &
$0.7 \pm0.2\,\,^{+0.3}_{-0.1} \pm 0.4 $   &
$308 \pm 12\,\,^{+15}_{-25}\,\,^{+66}_{-6}$ &
$0.27 \pm 0.15 \,\,^{+0.24}_{-0.09}\,\,^{+0.28}_{-0.14}$ \\
$\overline{K}^0 f_0(1370)$ $\times B(f_0(1370) \to \pi^+ \pi^-)$  & $1.8 \pm 0.1\,\,^{+0.2}_{-0.1}\,\,^{+0.2}_{-0.6}$     &
$85 \pm 4\,\,^{+4}_{-1}\,\,^{+34}_{-13}$  & $9.9 \pm 1.1 \,\,^{+2.4}_{-1.1}\,\,^{+1.4}_{-4.3}$ \\
$K^\ast_0(1430)^- \pi^+$ $\times B(K^\ast_0(1430)^- \to \overline{K}^0 \pi^-)$ & $2.0\pm 0.1\,\,^{+0.1}_{-0.2}\,\,^{+0.5}_{-0.1}$    &
$3 \pm 4 \pm 4 ^{+7}_{-15}$ &
$7.3\pm0.7\,\,^{+0.4}_{-0.9}\,\,^{+3.1}_{-0.7}$ \\
$K^\ast_2(1430)^- \pi^+$ $\times B(K^\ast_2(1430)^- \to \overline{K}^0
\pi^-)$ & $1.0\pm 0.1 \pm 0.1 \,\,^{+0.3}_{-0.1}$    &
$335\pm7\,\,^{+1}_{-4}\,\,^{+7}_{-24}$     &
$1.1 \pm 0.2 \,\,^{+0.3}_{-0.1} \,\,^{+0.6}_{-0.3}$ \\
$K^\ast(1680)^- \pi^+$ $\times B(K^\ast(1680)^- \to \overline{K}^0
\pi^-)$ &  $ 5.6\pm0.6\,\,^{+0.7}_{-0.4}\pm 4.0$ &
$174\pm 6\,\,^{+10}_{-3}\,\,^{+13}_{-19}$  &
$2.2\pm0.4\,\,^{+0.5}_{-0.3}\,\,^{+1.7}_{-1.5}$ \\
$\overline{K}^0 \pi^+ \pi^-$ non-resonant    & $1.1 \pm 0.3\,\,^{+0.5}_{-0.2}\,\,^{+0.9}_{-0.7}$ &
$340 \pm 11\,\,^{+30}_{-18}\,\,^{+55}_{-52}$ &
$0.9 \pm 0.4\,\,^{+1.0}_{-0.3}\,\,^{+1.7}_{-0.2}$
\end{tabular} 
\end{table} 
\end{landscape}
\end{document}